\documentclass[aps,prb,twocolumn,showpacs,showkeys,groupedaddress,amsmath,amssymb]{revtex4} 

\usepackage{natbib}
\usepackage{graphicx}
\usepackage[ansinew]{inputenc}
\usepackage{bm}
\usepackage{color}

\begin{document}

\title{Midgap states and band gap modification in defective graphene/h-BN heterostructures}

\author{B. Sachs}
\affiliation{I. Institut f{\"u}r Theoretische Physik, Universit{\"a}t Hamburg, Jungiusstra{\ss}e 9, D-20355 Hamburg, Germany}
\author{T. O. Wehling}
\affiliation{Institut f{\"u}r Theoretische Physik, Universit{\"a}t Bremen, Otto-Hahn-Allee 1, D-28359 Bremen, Germany}
\affiliation{Bremen Center for Computational Materials Science, Universit{\"a}t Bremen, Am Fallturm 1a, D-28359 Bremen, Germany}
\author{M. I. Katsnelson}
\affiliation{Radboud University of Nijmegen, Institute for Molecules and Materials, Heijendaalseweg 135, 6525 AJ Nijmegen, The Netherlands}
\affiliation{Theoretical Physics and Applied Mathematics Department, Ural Federal University, Mira Str.19, 620002 Ekaterinburg, Russia}
\author{A. I. Lichtenstein}
\affiliation{I. Institut f{\"u}r Theoretische Physik, Universit{\"a}t Hamburg, Jungiusstra{\ss}e 9, D-20355 Hamburg, Germany}
\affiliation{Theoretical Physics and Applied Mathematics Department, Ural Federal University, Mira Str.19, 620002 Ekaterinburg, Russia}

\begin{abstract}
The role of defects in van der Waals heterostructures made of graphene and hexagonal boron nitride (h-BN) is studied by a combination of ab initio and model calculations. Despite the weak van der Waals interaction between layers, defects residing in h-BN, such as carbon impurities and antisite defects, reveal a hybridization with graphene p$_{\rm z}$ states, leading to midgap state formation. The induced midgap states modify the transport properties of graphene and can be reproduced by means of a simple effective tight-binding model. In contrast to carbon defects, it is found that oxygen defects do not strongly hybridize with graphene's low-energy states. Instead, oxygen drastically modifies the band gap of graphene, which emerges in a commensurate stacking on h-BN lattices.
\end{abstract}
\maketitle

\normalsize

In recent years, van der Waals heterostructures \cite{grigorieva2013van}, i.e., stacks of two-dimensional crystals \cite{art:2dimcrystals} such as graphene and hexagonal boron nitride (h-BN), have attracted great research interest
 \cite{Slotman2015,Liu2016,defects2016,transport2016,interaction2016,Falko2016,Peeters2014}. 
These ``materials with tailored properties" \cite{novoselov2012two} exhibit novel phenomena, and, simultaneously, pave the way towards nanoelectronic applications such as tunneling transistors \cite{britnell2012field,georgiou2012vertical} or photovoltaic devices \cite{britnell2013strong,sachs2013doping,zhang2014ultrahigh}. Theoretically, however, many microscopic details behind the electronic and transport properties of these materials have not been understood so far. In particular, the role of defects in van der Waals heteros has rarely been studied, with few exceptions \cite{sachs2013doping,rodriguez2014ground,Ding_vacBN}.

In this paper, we study the effect of defective h-BN on graphene by combining first-principles density functional theory simulations with effective tight-binding (TB) models and Boltzmann transport theory. In particular, we demonstrate that impurities in h-BN enable the formation of midgap states in graphene. Midgap states are potentially responsible for vertical carrier tunneling in graphene/h-BN/graphene hybrids under zero electric field \cite{britnell2012field}. Here, we consider realistic carbon and oxygen defects as well as antisite defects in h-BN.

\begin{figure}
\includegraphics[width=0.99\linewidth]{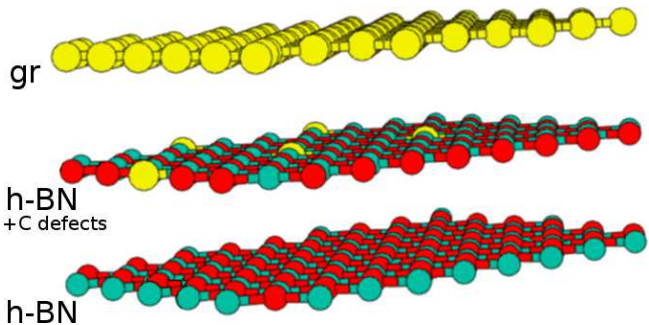}
\caption{(color online).Sketch of the graphene/h-BN heterostructure with carbon defects included in the topmost h-BN layer.}
\label{fig:struct}
\end{figure}

Our technical setup was chosen as follows: A supercell was constructed with the lateral dimension of 3x3 primitive h-BN unit cells, which comprised two h-BN layers covered with graphene, and contained impurities located in the topmost h-BN layer (Fig. \ref{fig:struct}). The substitutional carbon and oxygen atoms were either placed in a nitrogen or a boron site. Furthermore, we considered boron and nitrogen antisite defects. In the simulations, the lattice mismatch between graphene and h-BN of 1.8\% was neglected, thus, a commensurate stacking of lattices was assumed. Such a stacking is realistic in graphene/h-BN moir\'es with small rotation angles, which were recently shown to favor large regions in a commensurate state \cite{woods2014commensurate}. The stacking configuration of graphene on h-BN was chosen such that the B and N atoms were eclipsed by a carbon atom. The two h-BN sheets were stacked in A-A' order (eclipsed with B over N) analogous to the stacking of layers in bulk h-BN \cite{pease1950crystal} . Recent phonon calculation confirm the stability of similar heterostructures \cite{Slotman2014}.
While the bottom h-BN layer puts the model closer to the system of graphene on defective few layer or bulk hBN, it has no strong influence on the results and could be also neglected.

The DFT simulations were performed using the Vienna ab initio simulations package (VASP) \cite{kresse_vasp} with projector augmented (PAW) plane waves \cite{PAW1,PAW2}. The local density approximation (LDA) was employed to the exchange-correlation potential, which is superior to the generalized gradient approximation (GGA) for van der Waals solids and heterostructures \cite{LDAGGA_Schwarz,sachs2011hBN,hasegawa2004semiempirical}. The Brillouin zone was sampled by a 12x12x1 k-mesh, and a plane-wave cut-off of 500 eV was employed. All geometries were fully relaxed in order to find the optimized structure. It is predicted that carbon defects in isolated h-BN sheets induce magnetic moments \cite{azevedo2009electronic}. Since an exact many body solution for single defects structure should be paramagnetic and 
due to weak interlayer binding between graphene and h-BN, effects of local magnetic moments in h-BN are likely to be small. Therefore, spin-polarization was neglected in our simulations.

\begin{figure*}
\includegraphics[width=0.99\linewidth]{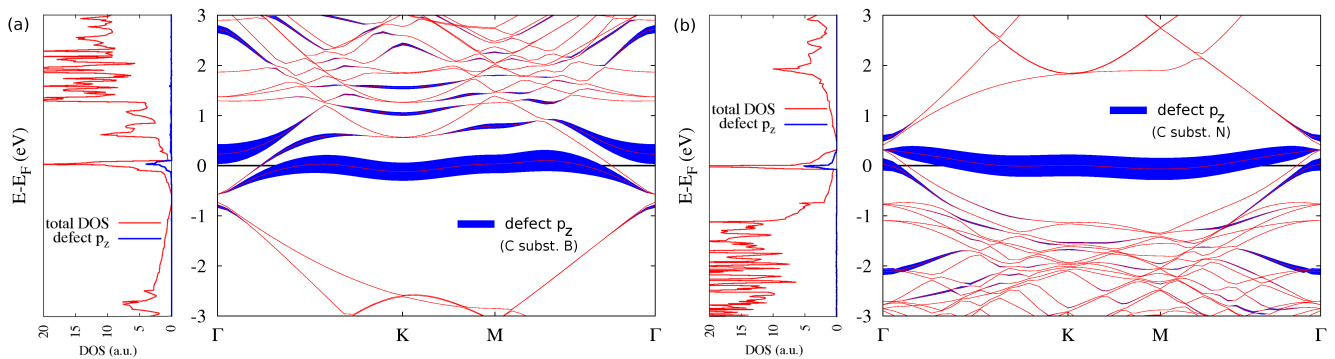}
\caption{(color online).Density of states (left panel) and band structure (right panel) of graphene on defective h-BN for the case of a carbon impurity substituting a boron atom (a) and carbon replacing a nitrogen atom (b). Due to the 3x3 supercell, Dirac bands are folded to the $\Gamma$ point. The $p_{\rm z}$ character of the impurity state is visualized by the thickness of blue-shaded regions in the band diagram and by the blue curve in the DOS plots.} 	
\label{fig:bands}
\end{figure*}

A reasonable indicator for the stability of defective heterostructures is the cohesive energy of the system which can be calculated as
\begin{equation}
E_{\rm coh}= \frac{n_{\rm B} E_{\rm B}+n_{\rm N} E_{\rm N}+n_{\rm C} E_{\rm C}+n_{\rm O} E_{\rm O}-E_{\rm tot}}{N_{\rm tot}},
\label{eq:cohesive}
\end{equation}
with $n_{\rm B}$, $n_{\rm N}$, $n_{\rm C}$, $n_{\rm O}$ the number of B, N, C, and O atoms in the unit cell, $E_{\rm B}, E_{\rm N}, E_{\rm C}, E_{\rm O}$ the respective total energies of the isolated atoms, $E_{\rm tot}$ the total energy of the defective (fully relaxed) heterostructure, and $N_{\rm tot}$ the total number of atoms in the unit cell (54 in our simulations). The energies are summarized in Tab. \ref{tab:cohesives}. From structural optimization, we find that the least stable defects, i.e., those with the lowest cohesive energies (antisites, O substituting B), lead to strong distortions of the surrounding B and N atoms with considerable corrugations of partly more than 1 \AA, which even slightly buckles the graphene sheet above the h-BN layer. Overall, the calculations indicate the highest stability of C defects, while antisite and O defects are less likely to occur.

\begin{table*}%
\centering
\begin{tabular}{|c|c|c|c|c|c|c|c|}
\hline
 & {\small pristine} &{\small  C subs. B} & {\small C subs. N} & {\small O subs. B} & {\small O subs. N} & {\small B antisite} & {\small N antisite}\\
\hline
{\small $\Delta E_{\rm coh}$ (meV)} & 0 & 20 & 70 & 180 & 70 & 160 & 80\\   
\hline
{\small gap (meV)} & 57 &  MG & MG  & 11 & 250&  MG& MG    \\
\hline
\end{tabular}
\caption{Cohesive energies relative to the cohesive energy of the pristine system ($E_{\rm coh, pristine}=9.648$ eV) together with the band gaps of the graphene/h-BN heterostructure (pristine, and with C, O, and antisite defects included with one impurity per 3x3 supercell). ``MG" indicates the presence of a midgap state.}
\label{tab:cohesives}
\end{table*}

The carbon defects in h-BN affect graphene's low energy states, as can be seen from the supercell density of states (DOS) and the band structure (Fig. \ref{fig:bands}). Two scenarios are depicted, carbon substituting a boron atom (a) and carbon replacing a nitrogen atom (b). The band structure reveals two features associated with the impurity: first, the Dirac cone, which can be discerned near the $\Gamma$ point (due to supercell-related Brillouin zone folding), is modified and shifted downwards in (a) by roughly 650 meV, while it is shifted upwards in (b) by 350 meV. With the carbon impurity, the h-BN layer contains either one additional electron (a) or a hole (b), and due to Fermi level pinning, the graphene sheet is likewise n-doped (a) or p-doped (b).

\begin{figure*}
\includegraphics[width=0.99\linewidth]{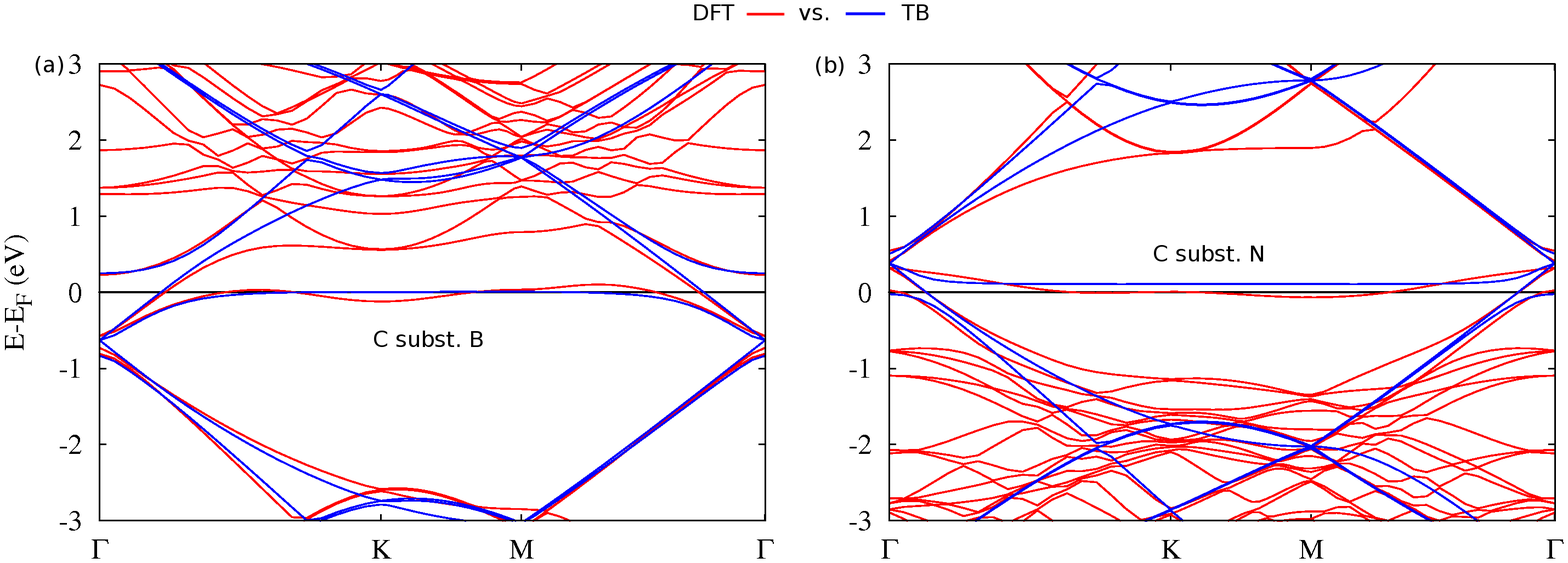}
\caption{(color online).Comparison of band structures as extracted from TB model calculations and DFT simulations: (a) A carbon impurity replaces a boron atom. (b) A carbon atom substitutes a nitrogen atom.} 	
\label{fig:bands_TB_DFT}
\end{figure*}

Second, the carbon defects do not only evoke a doping effect. Another modification of graphene states at the Dirac point can be observed as well that originates from the impurity states. This becomes clear from the position of defect-p$_{\rm z}$ states, which are visualized in Fig. \ref{fig:bands} through the thickness of blue ``fat bands" \cite{jepsen1995calculated}. One can discern a pronounced midgap state which is formed by a hybridization of the carbon impurity  and the graphene sheet. A similar effect is known to occur for many realistic adsorbates that are chemisorbed on graphene \cite{wehling2010resonant}; however, in the heterostructures investigated, here, the situation is different: 
a midgap state formation in graphene bands is unexpected since the impurity is located within the neighboring h-BN sheet at a relatively far distance of about 3.4 \AA. The bands in Fig. \ref{fig:bands} reveal that the hybridization with graphene depends on the position of the defect: the hybridization is more pronounced for a carbon defect substituting a boron atom. Compared to many weakly bound impurities \cite{wehling2008first,wehling2008molecular} (such as water molecules) on isolated graphene, the hybridization, here, is remarkably stronger. However, in comparison with adsorbates being covalently bound to isolated graphene \cite{wehling2010resonant}, hybridization is naturally weaker, which is supported by the nearly unmodified graphene/h-BN interlayer distance. The graphene atom above the impurity is only slightly shifted towards the h-BN layer by less than 0.01 \AA\ for both defect sites of the carbon atom.

Similarly, we find induced midgap states in the case of less stable B and N antisite defects. In particular, B antisite defects induce a large distortion of surrounding atoms in the h-BN layer as well as in the graphene sheet, which leads to drastic deformations of Dirac bands. Oxygen defects, in contrast, reveal only weak hybridization and no midgap states in graphene bands. However, we find a significant modification of the graphene band gap: while the gap of graphene on pristine h-BN in the considered stacking configuration is found to be 57 meV, it is strongly increased to 250 meV for an oxygen atom in a nitrogen site (oxygen in a boron site reduces the gap but is least stable). 
Obviously, the additional electron in the nitrogen site increases the modulation of the electrostatic potential, and, correspondingly, sublattice symmetry breaking is enhanced which increases the gap. Furthermore, the Dirac cone is down-shifted by about 850 meV, thus, graphene is strongly n-doped. 
This demonstrates the importance of impurities for the band gap formation in graphene on h-BN, which have been neglected so far in theoretical studies, while many-body effects \cite{song2013electron,jung2014origin,bokdam2014band} and moir\'e superlattice potentials \cite{sachs2011hBN,kindermann2012zero,mucha2013heterostructures,san2014electronic,van2014moire} have been widely investigated.

Next, we reconsider the midgap states of carbon impurities from a model point of view. Therefore, we set up a minimal low-energy tight-binding (TB) model of graphene electrons. Here, we take all states associated with h-BN and defects therein into account using the following effective Hamiltonian:
\begin{equation}
H = H_0 + V \left( a_{i'}^{\dagger}o + h.c. \right) + \left(\epsilon_{\rm imp} + \mu \right) o^{\dagger}o,
\label{eq:BNtb}
\end{equation}
where an impurity orbital $o$ with onsite energy $\epsilon_{\rm imp}$ couples to a graphene atom at a site $\vec{R}_{i'}$ via a hybridization $V$. $H_0$ is the nearest-neighbor tight-binding Hamiltonian of $p_{\rm z}$ electrons,
\begin{equation}
H_0 = -t \sum_{\left\langle i,j \right\rangle} \left( a_i^{\dagger} b_j + h.c. \right) + \mu \sum_i \left(a_i^{\dagger} a_i + b_i^{\dagger} b_i \right),
\label{eq:BNtbGr}
\end{equation}
with $a_i^{\dagger}$ ($b_i^{\dagger}$) the creation operator of a $p_{\rm z}$ electron acting on site $\vec{R}_i$ in sublattice A (B), and a hopping parameter $t$. The second term in both equations includes a chemical potential $\mu$ which is a constant term on the diagonal acting on all atoms, thus, a Fermi level shift. The parameters $V,$ $\epsilon_{\rm imp}$, $\mu$ and $t$ are \textit{a priori} unknown. We determine these in the following by connecting our effective TB model with the DFT simulations. The TB Hamiltonian provides a model-based insight into the defective heterostructure; moreover, determination of the model parameters from first-principles offers a quantitative picture of charge redistributions and allows for further processing in transport theory.

The effective TB model yields energy bands in good agreement with the low-energy states obtained from DFT (Fig. \ref{fig:bands_TB_DFT}). The model parameters $t$, $V$, $\epsilon_{\rm imp}$, and $\mu$ can be found from a fit to the DFT bands and are summarized in Tab. \ref{tab:TBparams}. The hybridization of $V \approx 0.9$ eV is remarkably high for an impurity which is not adsorbed on graphene, but resides in the neighboring h-BN layer. For comparison, methyl groups, which are strongly bound directly to graphene, have a hybridization of $V \approx 5$ eV \cite{wehling2010resonant}. Furthermore, the hybridization parameters confirm the DFT simulations of Fig. \ref{fig:bands}, which demonstrate a site-dependence of the hybridization and the doping effect.
\begin{table}%
\centering
\begin{tabular}{|c|c|c|c|c|}
\hline
& $t$ (eV)& $\mu$ (eV)& $ V $ (eV)& $\epsilon_{\rm imp}$ (eV)\\
\hline
C substituting B & +2.40& --0.63& 0.91& +0.70\\
C substituting N & +2.40& +0.38& 0.55& --0.28\\
\hline
\end{tabular}
\caption{Parameters used for the TB model (Eq. \ref{eq:BNtb}) as obtained by a fit to DFT bands. The employed boundary conditions were the same as in the DFT simulations (a 3x3 unit cell of graphene and one defect per unit cell). No sign for $V$ is given, because the resulting energy bands depend only on $|V|$.}
\label{tab:TBparams}
\end{table}
Considerable deviations between TB and DFT bands can only be found at higher energies far from the Dirac point due to the highly simplified model Hamiltonian. The TB model might easily be extended by including, e.g., the next-nearest neighbor hopping, or mass terms in order to model the intrinsic band gap of graphene on h-BN.

\begin{figure}
\includegraphics[width=0.99\linewidth]{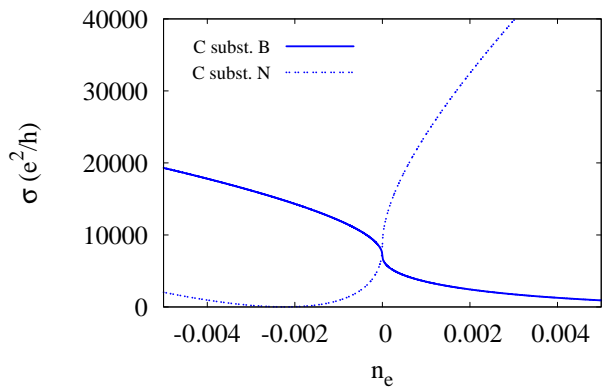}
\caption{(color online).Boltzmann conductivity versus the carrier concentration in graphene on defective h-BN for carbon impurities substituting boron (solid line) or nitrogen (dashed line) in an impurity concentration of $n_i=$ 0.1\%. A carrier concentration of $n_e=0.004$ corresponds to 1.54$\cdot 10^{13}$cm$^{-2}$.}
\label{fig:transport}
\end{figure}

Using our TB parameters one can in principle, calculate electron transport in the defect structure within the time-consuming Kubo formalism \cite{Yuan2012,Slotman2015,transport2016}. 
Here  we used a simple Boltzmann transport theory, to estimate the conductivity of graphene samples. For concentrations of $n_i$ defects per graphene carbon atom, we obtain within this framework
\begin{equation}
\sigma \approx \left( 2 e^2 / h \right) \left( 2 \pi n_i | T(E_{\rm F}) / D |^2 \right)^{-1},
\label{eq:trans}
\end{equation}
with $D=\sqrt{\sqrt{3} \pi} t \approx$ 5.6 eV and the scattering matrix $T(E_{\rm F})$, which depends on TB parameters $V$ and $\epsilon_{\rm imp}$. For more information about the Boltzmann approach and the $T$ matrix formalism, cf. Ref. \cite{Katsnelson_Graphene,wehling2010resonant,Hentschel2007,Peres2010,Couto2011} and references therein.
As shown in Ref. \onlinecite{wehling2010resonant}, the estimates of conductivities within the Boltzmann approach are qualitatively in line with full Kubo formula simulation at moderate impurity concentrations.
For a carbon defect concentration of $n_i=$ 0.1\% (corresponding to 3.8$\cdot 10^{12}$cm$^{-2}$), we obtain a conductivity as shown in Fig. \ref{fig:transport}. The conductivity exhibits a sublinear progression with the carrier concentration $n_e$ (which is determined by the Fermi level as $n_e=E_{\rm F}^2/D^2$). Although the defect concentration is considerable, a high conductivity is obtained around the neutrality point. It exceeds typically measured graphene/h-BN conductivities of $(10^2 - 10^3) e^2/h$ \cite{dean2010boron,sarma2011conductivity} by roughly one order of magnitude in both carbon scenarios. On the other hand, the conductivity has a distinct minimum ($\sigma_{\rm min} \approx 7 e^2/h$) at $n_e = -0.0022$ (corresponding to $-8.5\cdot 10^{12}$ cm$^{-2}$) for C substituting N. The same holds for C in a B site, but here, the minimum ($\sigma_{\rm min} \approx 39 e^2/h$) is located at a positive and very large concentration (around $4.5\cdot 10^{13}$cm$^{-2}$), which is outside the plot window. Overall, the curves indicate that defective h-BN layers are not the limiting factor for the graphene conductivity close to the neutrality point. Remember that the impurity concentration of $n_i = 0.1$\% is already large and $\sigma \sim n_i^{-1}$.  However, going away from the neutrality point ($n_e \neq 0$), defects in h-BN result in significant limitations of the conductivity even at much smaller impurity concentrations ($n_i \ll 0.1$\%) depending on the type of defects and the sign of $n_e$ . The results emphasize the importance of ab-initio-derived parameters in transport theory.

To conclude, we have demonstrated that realistic defects in h-BN strongly affect graphene electrons. Three important implications on electron transport should be mentioned: first, impurities such as carbon defects in h-BN lead to midgap states and, as a consequence \cite{wehling2010resonant}, to a scattering of carriers within the heterostructures in lateral direction. Second, the defects in h-BN might enable vertical transport in graphene/h-BN/graphene tunneling transistors in the absence of electric fields. Third, impurities such as oxygen atoms can significantly modify the graphene band gap. All scenarios must be carefully considered in the interpretation of transport experiments and deserve further investigation in the future.

Funding by the Deut\-sche For\-schungs\-ge\-mein\-schaft via SPP 1459 and the Graphene Flagship is acknowledged. MIK acknowledges funding from the European Union Horizon 2020 Programme under Grant No. 696656
Graphene Core. The work was supported by Act 211 Government of the Russian Federation, contract ? 02.A03.21.0006. Computations were performed at the North-German Supercomputing Alliance (HLRN) under Grant No. hhp00030.

\bibliography{refs}

\end{document}